\documentclass[runningheads]{llncs}
\usepackage[utf8]{inputenc}
\usepackage[T1]{fontenc}
\usepackage{graphicx}
\usepackage{amsmath}
\usepackage{cite}
\usepackage{xcolor}
\usepackage{pifont}
\usepackage{hyperref}
\usepackage{tabularx}
\usepackage{array}
\usepackage{rotating}
\usepackage{multirow, makecell}
\usepackage{footnote}
\usepackage{booktabs}
\makesavenoteenv{tabular}
\makesavenoteenv{table}

\newcommand{\ie}{\emph{i.e.},~}
\newcommand{\eg}{\emph{e.g.},~}
\newcommand{\ea}{\emph{et al.}~}
\newcommand{\vs}{\emph{vs.}~}

\author{Huy-Dung Nguyen \and
Michaël Clément \and
Boris Mansencal \and
Pierrick Coupé
}

\institute{Univ. Bordeaux, CNRS, Bordeaux INP, LaBRI, UMR 5800, 33400 Talence, France
\email{huy-dung.nguyen@u-bordeaux.com}}

\authorrunning{Nguyen \ea}

\title{3D Transformer based on deformable patch location for differential diagnosis between Alzheimer's disease and Frontotemporal dementia}

\titlerunning{3D Transformer based on deformable patch location for differential diagnosis}

\begin{document}

\maketitle

\begin{abstract}

Alzheimer's disease and Frontotemporal dementia are common types of neurodegenerative disorders that present overlapping clinical symptoms, making their differential diagnosis very challenging. Numerous efforts have been done for the diagnosis of each disease but the problem of multi-class differential diagnosis has not been actively explored. In recent years, transformer-based models have demonstrated remarkable success in various computer vision tasks. However, their use in disease diagnostic is uncommon due to the limited amount of 3D medical data given the large size of such models. In this paper, we present a novel 3D transformer-based architecture using a deformable patch location module to improve the differential diagnosis of Alzheimer's disease and Frontotemporal dementia. Moreover, to overcome the problem of data scarcity, we propose an efficient combination of various data augmentation techniques, adapted for training transformer-based models on 3D structural magnetic resonance imaging data. Finally, we propose to combine our transformer-based model with a traditional machine learning model using brain structure volumes to better exploit the available data. Our experiments demonstrate the effectiveness of the proposed approach, showing competitive results compared to state-of-the-art methods. Moreover, the deformable patch locations can be visualized, revealing the most relevant brain regions used to establish the diagnosis of each disease.

\end{abstract}

\section{Introduction}
Alzheimer's disease (AD) and Frontotemporal dementia (FTD) are the two most prevalent types of neurodegenerative disorders. They are the main cause of cognitive impairment and dementia \cite{bang_frontotemporal_2015}. Therefore, their differential diagnosis is crucial for determining appropriate interventions and treatment plans. However, these diseases share several overlapping symptoms such as memory loss and behavior changes, making their differential diagnosis challenging even when they have different clinical diagnostic criteria \cite{rascovsky_sensitivity_2011}. Indeed, several studies have demonstrated the limitations of cognitive tests in distinguishing patients with FTD from those with AD \cite{hutchinson_neuropsychological_2007, yew_lost_2013}. Furthermore, cognitively normal (CN) people may also exhibit some changes in behavior and memory as a result of the natural aging process. Consequently, an automatic tool for multi-class diagnosis (\ie AD \vs FTD \vs CN) is highly valuable in a real clinical context.

Several works have reported that AD and FTD are associated with brain structure atrophy \cite{pini2016brain, rosen2002patterns}, which can be visualized using structural magnetic resonance imaging (sMRI) \cite{du_different_2006, moller_alzheimer_2016}. This modality has been used to extract structure volumes \cite{du_different_2006} or used as input of convolutional neural networks (CNN) \cite{hu_deep_2021, nguyen2022deep} for differential diagnosis. In recent years, transformer-based models appear to be a promising alternative to CNN-based models in computer vision tasks. However, their application in disease diagnostic (\eg differential diagnosis) is still limited due to their computational demands and data requirements.

To alleviate computation problems, classification can be considered as a 2D problem. Lyu \ea and Jang \ea used 2D features extracted from MRI, both using a vision transformer (ViT) \cite{dosovitskiy2020image} for AD classification \cite{lyu2022classification, jang2022m3t}. However, the lack of spatial information in such 2D approaches may not be optimal. Regarding 3D methods, for AD diagnosis, Li \ea downsampled the input image before feeding it to their transformer \cite{li2022trans}, Zhang \ea reduced the feature map dimension by setting a big patch size for embedding \cite{zhang20223d}. However, these strategies may reduce the details of local regions. For natural image classification, other techniques to reduce computation are local attention \cite{liu2021swin} and deformable attention \cite{xia2022vision}. The idea of both methods is to reduce the size of the attention matrix by decreasing the number of query, key, and value points. In the case of deformable attention mechanism, key points can be visualized for better interpretation.

Transformer-based models are known to require a large amount of data to achieve high performance \cite{dosovitskiy2020image}. In medical imaging, the limited number of labeled sMRI makes it difficult to train these models effectively. In this situation, data augmentation plays an important role in the model generalization. While data augmentation has been shown to be effective for transformer in natural image classification \cite{touvron2021training}, its effectiveness in medical imaging has not been investigated.

In this paper, we first propose a 3D transformer-based architecture using a deformable patch location (DPL) module for the problem of multi-class differential diagnosis (\ie AD \vs FTD \vs CN). In the backbone, we employ local attention \cite{liu2021swin} instead of global one to reduce the computation. Our DPL module is inspired from the deformable attention\cite{xia2022vision}, however, unlike the original model, deformable points in DPL are determined for each sub-volume of the image rather than being shared across the entire image. Second, to alleviate data scarcity, we propose an efficient combination of various data augmentation techniques. The exploration of data augmentation for 3D transformer-based classification using sMRI has remained relatively unexplored until now, and our strategy aims to fill this gap. Moreover, our data augmentation allows a multi-scale prediction, improving our model performance. Finally, we propose to combine our transformer-based method with a support vector machine (SVM) using structure volumes to even better exploit the limited training data. As a result, our framework shows competitive results compared to state-of-the-art methods for multi-class differential diagnosis.

\section{Materials and method}

\subsection{Datasets and preprocessing}

Table~\ref{table:summary_participants} describes the number of participants used in this study. The data consisted of 3319 subjects from multiple studies: the Alzheimer's Disease Neuroimaging Initiative (ADNI) \cite{jack_alzheimers_2008}, the Frontotemporal lobar Degeneration Neuroimaging Initiative (NIFD) \footnote{Available at \url{https://ida.loni.usc.edu/}} and the National Alzheimer’s Coordinating Center (NACC) \cite{beekly_national_2007}. We only used T1-weighted MRIs at the baseline acquired with 3 Tesla machines. For the NIFD dataset, we only selected the behavior variant, progressive non-fluent aphasia, and semantic variant  sub-types.
The ADNI2 and NIFD datasets constituted our in-domain dataset while the NACC constituted our out-of-domain dataset. The in-domain dataset was used to perform a 10-fold cross-validation. The out-of-domain was used as an external dataset for evaluating the generalization capacity of the trained models.

\label{subsection:preprocessing}
The T1w MRI was preprocessed in 5 steps, which included (1) denoising \cite{manjon_adaptive_2010}, (2) inhomogeneity correction \cite{tustison_n4itk_2010}, (3) affine registration into MNI152 space ($181\times217\times181$ voxels at $1mm\times1mm\times1mm$) \cite{avants_reproducible_2011}, (4) intensity standardization \cite{manjon_robust_2008} and (5) intracranial cavity (ICC) extraction \cite{manjon_nonlocal_2014}. After that, we cropped at the image center a volume of size $144\times168\times144$ voxels to remove empty spaces. The brain structure volumes (\ie normalized volume in \% of ICC) were measured using a brain segmentation predicted by AssemblyNet \cite{coupe_assemblynet_2020}. These volume features were used as input for our SVM.
\begin{table}[t]
\caption{Number of participants.}\label{table:summary_participants}
\centering
\begin{tabular*}{0.7\textwidth}{@{\extracolsep{\fill}}llccc}
\toprule
& Dataset & CN & AD & FTD \\

\midrule
\multirow{2}{*}{In-domain} & ADNI2 &  180 & 149 & \\

\cmidrule{2-5}
& NIFD & 136 & & 150\\

\midrule
Out-of-domain& NACC & 2182 & 485 & 37\\

\bottomrule
\end{tabular*}
\end{table}

\subsection{Method}
\label{section:method}
\begin{figure}[ht]
    \centering
    \includegraphics[width=0.75\textwidth]{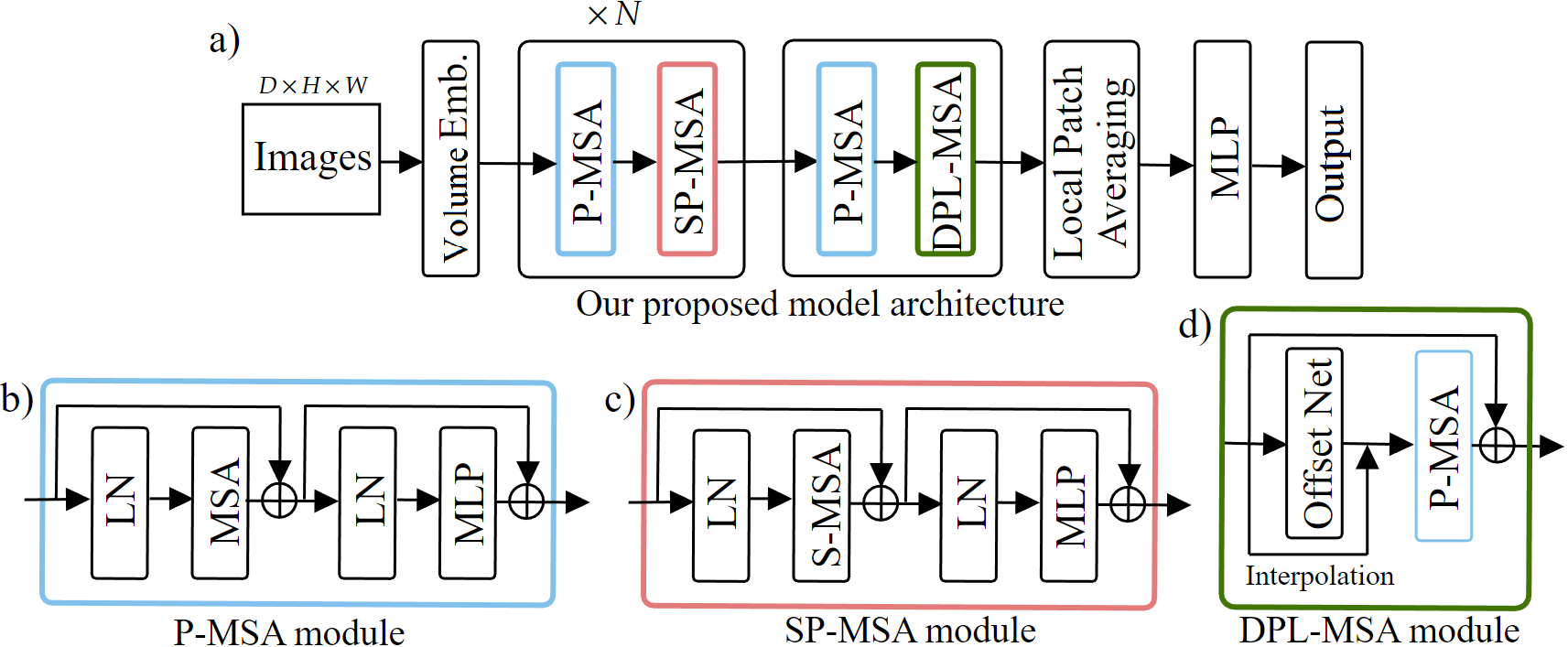}
    \caption{The architecture of our proposed model}
    \label{fig:model_architecture}
\end{figure}
\textbf{Overview} Figure \ref{fig:model_architecture} shows an overview of our proposed model. Our model is composed of four parts: a volume embedding (VE), $N$ blocks of a patch multi-head self-attention (P-MSA) followed by a shift patch multi-head self-attention (SP-MSA - the main building block of Swin \cite{liu2021swin}), a deformable patch location multi-head self-attention module (DPL-MSA) and a local patch averaging layer followed by a multi-layer perceptron (MLP). Intuitively, the VE module encodes an MRI to a 3D volume of tokens. The $N$ blocks of P-MSA and SP-MSA process these tokens as a attention-based feature extractor. Then, the DPL-MSA block predicts deformable patch locations and performs attention on them. While standard transformer-based approaches perform a global average of all the patches together \cite{liu2021swin}, in our method we perform local average of patches in the same area (\ie sub-volume). To this end, we divide the brain feature map into 27 sub-volumes ($3 \times 3 \times 3$ areas evenly distributed along 3 axis). This is because different brain locations may be affected by a disease differently, thus should be weighted differently in the model decision. Finally, we use an MLP for classification.

\noindent
\textbf{Volume embedding} We start with a preprocessed image of size $144 \times 168 \times 144$ (at $1mm^3$) (see Figure \ref{fig:model_architecture}a). The VE module uses a CNN (similar to \cite{touvron2022three}) to embed the input into token vectors (with an embedding dimension of 96). This results in a 96-channel 3D feature map of size $36 \times 42 \times 36$.

\noindent
\textbf{Feature extractor} The obtained 3D feature map is fed into three (P-MSA + SP-MSA) blocks. The details of each block are presented in \ref{fig:model_architecture}b,c. Our implementation of these blocks is based on \cite{liu2021swin}. The local attention size is set to $6 \times 7 \times 6$. By using attention mechanism, the size of feature maps remain unchanged.

\noindent
\textbf{DPL block} Taking the output of the feature extractor, we first update the feature map with a P-MSA module (see \ref{fig:model_architecture}a). We then split it into $6 \times 6 \times 6$ reference patches of size $(p_x, p_y, p_z) = (6 \times 7 \times 6)$. Their centers are denoted as: $(x_{ct}^{i}, y_{ct}^{i}, z_{ct}^{i})$. The coordinates of these points are normalized in $[0,1]$. Each reference patch is used as input of an offset network (see Figure \ref{fig:model_architecture}d) to predict the offset logits $(\delta_x^{i}, \delta_y^{i}, \delta_z^{i})$. The deformable patch center $(x_{Dct}^{i}, y_{Dct}^{i}, z_{Dct}^{i})$ is then calculated by:
$x_{Dct}^{i} = x_{ct}^{i} + \tanh{\delta_x^{i}} / (2 \times p_x)$ (idem for $y_{Dct}^{i}$ and $z_{Dct}^{i}$). Based on the deformable patch centers, we interpolate our feature map to obtain the corresponding deformable patches of size $p_x \times p_y \times p_z$. After that, we apply a P-MSA module to these deformable patches. Finally, a shortcut from reference patches is added to the output of the P-MSA module (see Figure \ref{fig:model_architecture}d).

\noindent
\textbf{Local patch averaging} We consider the obtained 96-channel 3D brain feature map (of size $36 \times 42 \times 36$) as a $3 \times 3 \times 3$ areas of size $12 \times 14 \times 12$ voxels, which are evenly distributed along 3 dimensions. We first average each deformable patch to a 96-channel mean token (of size $1 \times 1 \times 1$). Then, all the mean tokens located in a same area are averaged. Finally, we concatenate the obtained tokens and feed it into a MLP for classification.

\subsection{Data augmentation}
\label{section:da}
In this part, we describe our combination of data augmentation techniques. We start with mixup, which has been known to reduce overfitting in various applications \cite{zhang_mixup_2018}. Following this, we apply a series of affine transformations, including rotation and scaling, commonly used in medical imaging applications \cite{hussain2017differential,nalepa2019data}. To further enhance our augmentation process, we randomly crop images at an arbitrary position (with a probability $p$) and resize them to match the input resolution. This technique, similar to "Random resized crop" in 2D imaging \cite{touvron2019fixing}, mitigates overfitting and allows evaluation at both global and local views of an image. During inference, we ensemble predictions from multiple views to improve the model performance. In Section \ref{subsection:ablation_study}, we demonstrate the importance of each of these techniques on our framework accuracy.

\subsection{Validation framework and ensembling}
When evaluating our models, we made two predictions for each image: one for the whole image and one for a crop of that image. The cropping position was selected from nine cropping positions: a center crop and eight crops at corners. For each trained model, the crop position that produced the lowest loss on the validation set was selected. Finally, we averaged the two obtained results.

To further exploit the limited amount of training data, we combined (\ie average) the transformer prediction with SVM prediction based on brain structures volumes (see Section \ref{section:implementation_details}).

\subsection{Implementation details}
\label{section:implementation_details}
The offset network consisted of 3 layers: 3D convolution with 24 channels, kernel = (6, 7, 6), GELU activation \cite{hendrycks2016gaussian} and another 3D convolution with 3 channels, kernel = 1. For data augmentation, rotation range was $\pm 0.05 rad$ and scale range was $[0.9, 1.1]$, the crop size was (132, 154, 132), the probability $p=0.7$. The model was trained for 300 epochs using AdamW optimizer \cite{loshchilov2017decoupled}, cosine learning rate scheduler (start at 3e-4 and end at 5e-5). To train the SVM models, we used a grid search of three kernels (linear, polynomial, and gaussian) and 50 values of the hyper-parameter C in $[10^{-2}, 10^2]$ on the validation for tuning hyper-parameters. The SVM models used the same train/validation/test (70\%/20\%/10\%) splits of in-domain data during cross-validation than our deep learning models.

\section{Experimental results}
In this study, we first performed a 10-fold cross-validation on in-domain dataset. This resulted in 20 models (10 Transformers and 10 SVM models). We concatenated the prediction of 10 test folds to compute the global in-domain performance. For out-of-domain evaluation, we averaged all 10 predictions to estimate the model performance. We used 3 metrics to assess the model performance: accuracy (ACC), balanced accuracy (BACC) and area under curve (AUC).
\subsection{Ablation study}
\label{subsection:ablation_study}
\textbf{Performance study}
In this part, we studied the impact of each contribution on our model performance. These factors could be organized into 4 groups: Input type (2D/3D), architecture (local patch averaging, non linear volume embedding), validation framework (multi-scale prediction) and ensemble (combination with SVM). The used data augmentation schema was described in \ref{section:da}. Table \ref{table:ablation_study_1} showed the results of the comparison.

First, we implemented a basic 2D transformer-based architecture (exp. 1) and its 3D version (exp. 2) to see if the spatial information from 3D input is valuable. We observed that the 3D version was better than the 2D version in all metrics. Second, using local patch averaging (exp. 3) improved our model performance, confirming the effectiveness of assigning different weights to different brain areas. Third, the nonlinear volume embedding (exp. 4) could also improve the performance of transformer, which was inline with \cite{touvron2022three}. Then, the DPL module demonstrated an improvement in performance across almost all metrics (exp. 5). Finally, the multi-scale prediction (exp. 6) and ensembling (exp. 7) increased even more our model performance in both in-domain and out-of-domain data.
\begin{table}[t]
\caption{Ablation study of the model performance. Results obtained using the data augmentation described in \ref{section:da}. Gray text, symbols: that option is the same as in the previous experiment.  {\color{red} Red}, {\color{blue} Blue}: best, second result.}\label{table:ablation_study_1}
\begin{center}
\begin{minipage}{\textwidth}
\begin{tabular*}{\textwidth}{@{\extracolsep{\fill}}ccccccccccccc}
\toprule
    \multirow{5}{*}{No.} & 
    \multirow{5}{*}{\rotatebox[origin=l]{90}{\parbox{18mm}{2D/3D}}} & 
    \multirow{5}{*}{\rotatebox[origin=l]{90}{\parbox{18mm}{Local patch\\averaging}}} &
    \multirow{5}{*}{\rotatebox[origin=l]{90}{\parbox{18mm}{Nonlinear VE}}} &
    \multirow{5}{*}{\rotatebox[origin=l]{90}{\parbox{18mm}{DPL module}}} &
    \multirow{5}{*}{\rotatebox[origin=l]{90}{\parbox{18mm}{Multi-scale\\prediction}}} &
    \multirow{5}{*}{\rotatebox[origin=l]{90}{\parbox{18mm}{Combination\\with SVM}}}\\
    \\
    \\
& & & & & & & \multicolumn{3}{c}{In-domain} & \multicolumn{3}{c}{Out-of-domain}\\
& & & & & & & ACC  & BACC  & AUC  & ACC  & BACC  & AUC \\

\midrule

1 & {\color{black!40}2D} & {\color{black!40}\ding{55}} & {\color{black!40}\ding{55}} & {\color{black!40}\ding{55}} & {\color{black!40}\ding{55}} & {\color{black!40}\ding{55}} & 68.8 & 64.1 & 81.1 & 77.4 & 63.3 & 78.4 \\
2 & 3D & {\color{black!40}\ding{55}} & {\color{black!40}\ding{55}} & {\color{black!40}\ding{55}} & {\color{black!40}\ding{55}} & {\color{black!40}\ding{55}} & 78.4 & 74.7 & 90.1 & 81.5 & 75.2 & 87.8\\
3 & {\color{black!40}3D} & \ding{51} & {\color{black!40}\ding{55}} & {\color{black!40}\ding{55}} & {\color{black!40}\ding{55}} & {\color{black!40}\ding{55}} & 82.9 & 79.5 & 92.7 & 85.4 & 78.2 & 89.3\\
4 & {\color{black!40}3D} & {\color{black!40}\ding{51}} & \ding{51} & {\color{black!40}\ding{55}} & {\color{black!40}\ding{55}} & {\color{black!40}\ding{55}} & 83.6 & 80.3 & 92.5 & 86.6 & 79.7 & 89.9\\
5 & {\color{black!40}3D} & {\color{black!40}\ding{51}} & {\color{black!40}\ding{51}} & \ding{51} & {\color{black!40}\ding{55}} & {\color{black!40}\ding{55}} & 83.4 & 80.7 & 93.4 & 87.1 & 80.1 & 90.5\\
6 & {\color{black!40}3D} & {\color{black!40}\ding{51}} & {\color{black!40}\ding{51}} & {\color{black!40}\ding{51}} & \ding{51} & {\color{black!40}\ding{55}} & {\color{blue}85.2} & {\color{blue}82.5} & {\color{blue}94.1} & {\color{blue}87.7} & {\color{blue}80.7} & {\color{blue}91.0}\\

\midrule
7 & {\color{black!40}3D} & {\color{black!40}\ding{51}} & {\color{black!40}\ding{51}} & {\color{black!40}\ding{51}} & {\color{black!40}\ding{51}} & \ding{51} & {\color{red}86.2} & {\color{red}83.4} & {\color{red}94.5} & {\color{red}89.3} & {\color{red}82.8} & {\color{red}91.6}\\

\bottomrule
\end{tabular*}
\end{minipage}
\end{center}
\end{table}

\noindent
\textbf{Data augmentation study}
Table \ref{table:ablation_study_2} shows the contribution of each data augmentation technique to our model performance. The ensembling with SVM was removed for analysis and the multi-scale evaluation was applied only when multi-crop was used. First, without any data augmentation, the obtained result (exp. 1) was lower than in other experiments. Second, combining different augmentations (exp. 2, 3, 4) progressively improved the model's generalization. This showed the effectiveness of our data augmentation for medical imaging applications.

\begin{table}[t]
\caption{Ablation study of the data augmentation. Gray symbols: that option is the same as in the previous experiment.  {\color{red} Red}, {\color{blue} Blue}: best, second result.}\label{table:ablation_study_2}
\begin{center}
\begin{minipage}{\textwidth}
\begin{tabular*}{\textwidth}{@{\extracolsep{\fill}}cccccccccc}
\toprule
\multirow{3}{*}{No.} & \multirow{3}{*}{\rotatebox[origin=l]{90}{Mixup}} & \multirow{3}{*}{\rotatebox[origin=l]{90}{\parbox{9mm}{Rand.\\affine}}} & \multirow{3}{*}{\rotatebox[origin=l]{90}{\parbox{9mm}{Multi\\crops}}} \\
& & & &\multicolumn{3}{c}{In-domain} & \multicolumn{3}{c}{Out-of-domain}\\
& & & & ACC & BACC & AUC & ACC & BACC & AUC \\

\midrule

1 & {\color{black!40}\ding{55}} & {\color{black!40}\ding{55}} & {\color{black!40}\ding{55}} & 74.6 & 69.0 & 87.8 & 84.3 & 73.3 & 87.3 \\
2 & \ding{51} & {\color{black!40}\ding{55}} & {\color{black!40}\ding{55}} & 77.6 & 72.0 & 88.4 & 84.8 & 76.0 & 87.4\\
3 & {\color{black!40}\ding{51}} & \ding{51} & {\color{black!40}\ding{55}} & {\color{blue}82.1} & {\color{blue}78.9} & {\color{blue}91.5} & {\color{blue}86.2} & {\color{blue}78.6} & {\color{blue}90.0}\\

\midrule
4 & {\color{black!40}\ding{51}} & {\color{black!40}\ding{51}} & \ding{51} & {\color{red}85.2} & {\color{red}82.5} & {\color{red}94.1} & {\color{red}87.7} & {\color{red}80.7} & {\color{red}91.0}\\

\bottomrule
\end{tabular*}
\end{minipage}
\end{center}
\end{table}

\subsection{Comparison with state-of-the-art methods}
In this section, we compare our results with current state-of-the-art methods for the multi-class diagnosis AD \vs FTD \vs CN. Hu \ea proposed an CNN-based architecture inspired by Resnet which processes the whole 3D MRI for classification~\cite{hu_deep_2021}. Ma \ea used a MLP with cortical thickness (Cth) and brain structure volumes extracted from a 3D MRI \cite{ma_differential_2020}. They also used a generative adversarial network to generate new data to prevent over-fitting. More recently, Nguyen \ea used a large number of CNN to grade brain regions. The grading values were then averaged for each brain structure and used as input of a MLP for classification \cite{nguyen2022deep}. For a fair comparison, we reimplemented these methods and trained them under the same training setting as our method and on the same data. Table \ref{table:sota_3_classes} shows the results of the comparison.

Overall, our method presented most of the time the best performance in all metrics (\ie ACC, BACC and AUC) and for both in-domain and out-of-domain data. Moreover, our method was the only method based on the transformer mechanism. This suggested that transformer-based methods can obtain competitive results compared to CNN-based networks even with a limited amount of data.
\begin{table}[ht]
    \centering
    \caption{Comparison with state-of-the-art methods.  {\color{red} Red}, {\color{blue} Blue}: best, second result.}
    \label{table:sota_3_classes}
    \begin{center}
\begin{minipage}{\textwidth}
\begin{tabular*}{\textwidth}{@{\extracolsep{\fill}}lcccccc}
\toprule
\multirow{2}{*}{Method} & \multicolumn{3}{c}{In-domain} & \multicolumn{3}{c}{Out-of-domain}\\
 & ACC & BACC & AUC & ACC & BACC & AUC \\

\midrule

CNN on intensities \cite{hu_deep_2021}
& 76.3 & 72.5 & 90.0 & 85.2 & 68.8 & {\color{blue} 86.5}  \\
MLP on Cth and volumes \cite{ma_differential_2020}      
& 77.1 & 75.9 & 86.4 & 69.1 & 74.6 & 87.5  \\
3D Grading \cite{nguyen2022deep} 
& {\color{blue} 86.0} & {\color{red} 84.7} & {\color{blue} 93.8} & {\color{blue} 87.1} & {\color{blue} 81.6} & {\color{red} 91.6}\\

\midrule
Our method & {\color{red}86.2} & {\color{blue}83.4} & {\color{red}94.5} & {\color{red}89.3} & {\color{red}82.8} & {\color{red}91.6} \\

\bottomrule
\end{tabular*}
\end{minipage}
\end{center}
\end{table}

\subsection{Visualization of deformable patch location}
\begin{figure}[t]
    \centering
    \includegraphics[width=0.6\textwidth]{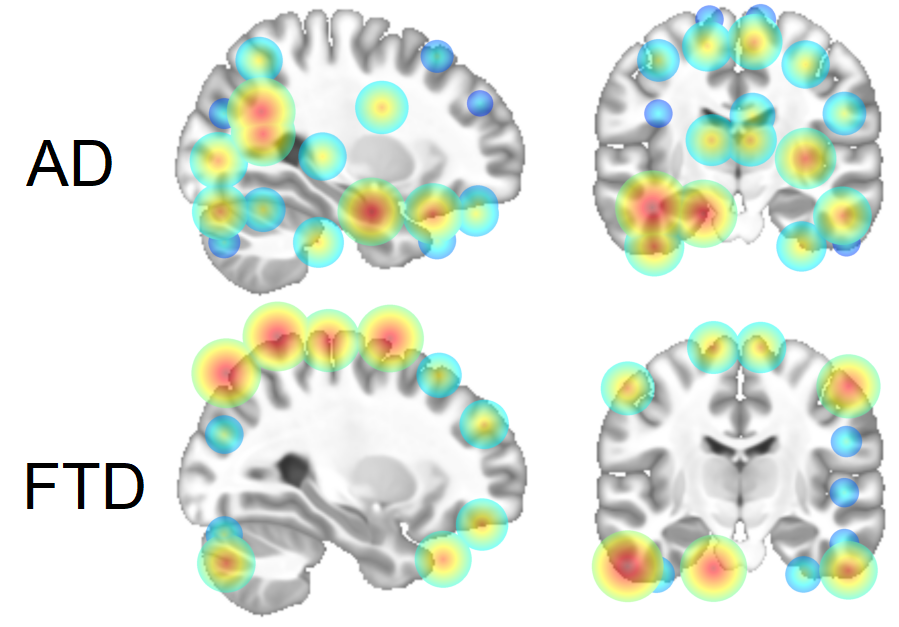}
    \caption{Visualization of deformable patch locations. The importance of each patch was estimated with GradCAM. Warmer color, larger radius mean higher importance score.}
    \label{fig:visu}
\end{figure}

Figure \ref{fig:visu} shows the centers of deformable patch locations for patients with AD and FTD. For each patient group, the patch center positions are calculated as the averaged center locations from our ten models. To enrich visual comprehension, we utilized GradCAM to attribute an importance score within the range of $[0, 1]$ to each patch. Patches obtaining an importance score above 0.3 are displayed. Furthermore, a higher importance score is visually represented by a larger circle, and the warmth of the circle's color increases with the score.

The obtained results were coherent with the current knowledge about these diseases. Indeed, for AD patients, the structures that obtained higher score were the left hippocampus \cite{schuff_mri_2009}, bilateral entorhinal cortex, bilateral ventricle \cite{coupe_lifespan_2019} and parietal lobe \cite{silhan2020parietal}. In FTD patients, the frontal pole\cite{boeve_advances_2022}, superior frontal gyrus \cite{brambati2007tensor} and left temporal cortex \cite{whitwell_distinct_2009} were highlighted.

\section{Conclusion}
Our study presents a novel 3D transformer model, which incorporates a deformable patch location module for the differential diagnosis between cognitively normal subjects, patients with Alzheimer's disease and patients with Frontotemporal dementia. The proposed module enhances the model's accuracy and provides useful visualizations that reveal insights into each disease. To address the problem of limited training data, we designed a combination common data augmentations for training transformer models using 3D MRI. Furthermore, we proposed to combine both our deep learning model and an SVM using brain structure volumes to even better exploit the limited data. As a result, our framework showed competitive performance compared to state-of-the-art methods.

\newpage

\appendix

\bibliographystyle{splncs04}
\bibliography{MLMI-2023-Paper_20}

\end{document}